# COOK'S THEORY AND TWENTIETH CENTURY MATHEMATICS

Li Chen

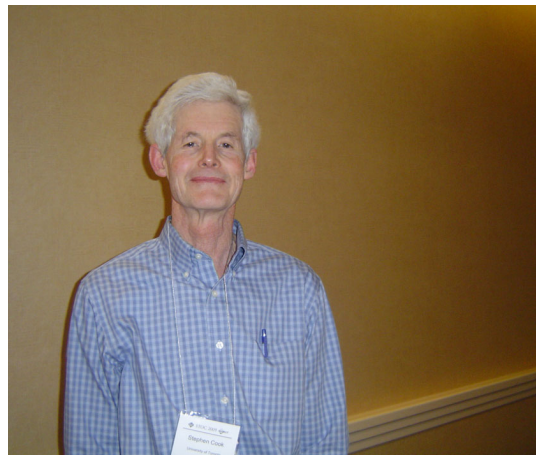

Today, many top mathematicians know about or even study computational complexity. As a result, mathematicians now pay more attention and respect to the foundation of computer science and mathematics. The foundation of mathematics, which is highly related to logic and philosophy, was for a longest time ignored by a considerable number of mathematicians. Steve Smale may have been the first pure mathematical figure who conducted research on complexity theory; he published his first complexity paper in 1985. Currently, Terry Tao is the most active mathematician doing research in this area.

When Peter Lax, who may be the most renowned mathematician today, was recently asked about the development of mathematics, he mentioned that complexity might play a leading role. However, he also questioned active researchers of complexity why they love to search for something difficult while we (as mathematicians) want to find something simple. Charlie Fefferman chose a model of complexity that accounts for research in Whitney's problem in mathematical analysis, which is highly related to the model used in computer science. (Among the four mathematicians mentioned above, three are Fields Medallists and one received the Wolf and Abel Prizes.)

In the history of computational complexity, it is widely accepted that Stephen Arthur Cook, a professor at the University of Toronto, has made the most important contributions.

What is the theory of complexity? Complexity deals with the time and space needs of solving one or a category of problems in a specific computing model. We usually use a Turing machine for computation. At the core of computational complexity is the NP-Completeness theory, of which the fundamental theorem is Cook's theorem.

However, the general public found it difficult to understand Cook's theorem and NP-Complete theory. A large number of computer scientists are unable to give a full explanation of Cook's theorem because they do not know what "NP" is. This case is worse than the case of the Brouwer fixed point theorem in mathematics, which 90% of mathematicians know about but 99% do not know how to prove.

The modern computational complexity started with Michael O. Rabin's "degree of difficulty of computing" for functions published in 1960. Hartmanis and Stearns then invented the concept of computational complexity in algorithms in 1965. Edmonds first discussed polynomial time algorithms in his paper related to graphs. In 1969, Hopcroft and Ullman published a fundamentally important book in theoretical computer science titled Formal languages and their relation to automata. The book presents most of the concepts and also includes M. Blum's axioms on complexity.

Then Cook came to the center of the "play." In 1971, Cook published a paper titled "The complexity of theorem-proving procedures" in the Proceedings of the 3rd Annual ACM Symposium on Theory of Computing, pp. 151–158, 1971. This was just one year after UC Berkley denied his tenure and the same year that he relocated to the University of Toronto as an Associate Professor.

Wikipedia has a detailed description of his biography: "Cook received his Bachelor's degree in 1961 from the University of Michigan, and his Master's degree and Ph.D. from Harvard University, respectively in 1962 and 1966. He joined the University of California, Berkeley, mathematics department in 1966 as an Assistant Professor, and stayed there till 1970 when he was infamously denied tenure. In a speech celebrating the 30th anniversary of the Berkeley EECS department, fellow Turing Award winner and Berkeley professor Richard Karp said that, 'It is to our everlasting shame that we were unable to persuade the math department to give him tenure.' Cook joined the faculty of University of Toronto, Computer Science and Mathematics Departments in 1970 as an Associate Professor, where he was promoted to Professor in 1975 and University Professor in 1985."

Understanding Cook's paper is not difficult today. However, not everyone in complexity has really read or studied his seminal work. It is definitely astonishing how Cook came up with the idea of polynomial time reduction and applied it to every case regarding a non-deterministic Turing machine initiated by Rabin and Scott in 1959.

Cook did not think that his treatment was very difficult since "the idea of reduction has been used in mathematical logic." In fact, Turing defined the computability class by introducing the concept of reduction in 1939 (now called Turing reduction). Cook was a PhD student of Wang Hao, a notable mathematical logician at Harvard and the concept of reduction seemed quite natural for Cook, who stated that: "I rushed an abstract to meet the deadline of the conference." Then he said, "After the paper was accepted, I revised it and sent it out." He insisted that he did not spend a lot of time thinking about the problem before completing his paper. Cook added: "In fact I did not have the idea when I submitted my paper (the abstract)--- which accounts for the rather strange title 'The Complexity of Theorem Proving Procedures.' I got the NP-completeness idea after my submission was accepted, and incorporated it in the version in the conference proceedings." (Cook was quiet, very much like a Kung Fu grand master from the movies: nice, proud, and full of confidence. If you have seen *Kung Fu Panda*, you will remember that "the best ingredient is no ingredient." The road for being a grand master is different in each case.)

What did Cook really accomplish in the eight pages? He proved that the SAT problem is NP-Complete. This resulted in what we know today as NP-Completeness theory. In fact, Cook created a tool called the polynomial time reduction. This tool still dominates today in the theory of complexity.

It is fairly simple to explain what polynomial time reduction is. Reduction means that in order to identify a set, it can be transferred to decide another set. The need of polynomial time reduction was the key idea behind Cook's theories. Obviously we cannot know exactly how he came up with these ideas (and how to arrange a proof) at that time, but we can go over a step-by-step explanation of Cook's theorem.

Let us take a real life example. Today, we are in an economic crisis and there are many ways to solve the problem. The most direct way is to give money to every family. However, this method would be too costly and uncontrollable. We could give money to the banks instead and then the banking system can bring the economy back up. However, we would have to restrict the banking system by only allowing them to do what the government wants, restricting their activities by a subset. The banker would say that we have two types of money: (a) money that transfers existing the specific purpose. Transferring money, and (b) money used to solve the problem. Solving money. The total cost of solving such economic crisis is smaller than or equal to the sum of Transferring money and Solving money. This is due to the existence of other ways of solving the economic problem such as putting the money into constructing railway systems, schools, and public services.

*In the abstract of his paper, Cook writes* "It is shown that any recognition problem solved by a polynomial time-bounded nondeterministic Turing machine can be 'reduced' to the problem of determining whether a given prepositional formula is tautology. Here, 'reduce' means, roughly speaking, that the first problem can be solved deterministically in polynomial time provided an oracle is available for solving the second."

Let us first explain the second part of the paragraph, i.e. what "reduce" means in his original paper. He means that the first problem can be solved in polynomial time if we know the solution to the second problem. The oracle here means that it can just tell us the solution.

This is equivalent to saying that only polynomial time is allowed to transfer one system to another. However, most of the money must be used in real problem solving. By the way, when the order of a polynomial is high, practically, the cost is also very high.

Now we come to the first part of his statement. The "recognition problem" is in regard to the decision problem, which only has a yes or no answer. For example, does a graph contain a cycle that visits each node only once (The Hamiltonian Cycle)? Also, what is a Turing machine? It was the model computer invented by Turing in 1936. It used tape and contained a head that could write and erase data on the tape. "Deterministic" means that the machine can choose an action at any given time. A non-deterministic Turing machine can choose two or more actions, for instance, whether to go forwards or backwards. Cook focuses on whether "there is a way to the end (we either accept the input or not)." In addition, a prepositional formula refers to a Boolean expression and tautology means that every case is true.

If a set of strings is said to be accepted by a Turing machine, we mean that for any string in the set there is a path/sequence of actions from beginning to end and at the end, the Turing machine says "yes." Each action step would form an assertion ("we did it here!") and all assertions in the path would be true. The union of all possible paths would determine whether the machine could accept the string.

This explanation now comes to an end because every decision-making problem that can be solved in polynomial time by the non-deterministic Turing machine (NP problems) is equivalent to the Tautology Problem of Boolean expression (in disjunctive normal form).

This original version of NP-completeness was a little different from the satisfiability problem in the conjunctive normal form due to Karp's simplification of P-reduction in 1972. Tautology of the union is equivalent to the complement of there being a true path, which is what we call the SAT problem today.

We now have the first hardest problem in NP and many problems can be proven to have the same difficulty level. The greatness of Cook method is that it places all problems in one unified category. It does not try to solve a specific problem, much like many of the problems we encounter today. This is why we can say that Cook's theorem reflects the great beauty of mathematics.

Cook's theorem explains why we are not able to find efficient algorithms for many important problems. However, such difficulties have also had positive outputs. It provided a more hurdles for decryption. Another direct application of Cook's theorem is public key cryptography. The Diffie–Hellman protocol was invented in 1976 and RSA was introduced in 1978. Without NP-completeness theory, we cannot imagine that public key cryptography would exist. This is especially true for the possible existence of the one-way trap door function. In fact, NP-completeness theory provided such a confidence to Diffie–Hellman's revolutionary public-key cryptosystem (see their paper *New Directions in Cryptography*). In the 1960-70s, scientists found uses for modern algebra in communication coding theory. On a same, or even higher, level of importance, Cook's theory is one of the foundations for the modern public key system since no one is likely to obtain a feasible method to solve an NP-Complete problem in the near future using practical algorithms. Cook's theory also provided a new thinking direction for solving a long-lasting math problem that involved computing the permanent of a matrix. In 1979, Lesllie Valiant proved that this problem is at least as hard as an NP-complete problem, meaning that there is no simple calculating procedure as we can obtain the determinate of a matrix using Gaussian elimination.

Why do we say that Cook's theory is one of the most important developments of 20$^{th}$ Century Mathematics?

It would be a very difficult task for people to review the major theories developed in the 20$^{th}$ Century since it is still too early to tell. Nevertheless, we can still list some of the most important theories to us currently. In fact, some authors have already evaluated these theories. For example, the science article writer John Casti listed ten great theories in his books called *Five Golden Rules* and *Five More Golden Rules*. These are: (1) the minimax theorem (game theory), (2) the Brouwer fixed-point theorem (topology), (3) Morse's theorem (singularity theory), (4) the halting theorem (theory of computation), (5) the simplex method (optimization theory), (6) the Alexander polynomial (knot theory), (7) the Hopf bifurcation theorem (dynamical system theory), (8) the Kalman filter (control theory), (9) the Hahn-Banach theorem (functional analysis), and (10) the Shannon coding theorem (information theory).

We do not know if mathematicians agree with him, but at least he had the courage to compile a list. It is believed that this list covers the major portions of the most important developments in mathematics.

At the beginning of the 20$^{th}$ Century, Mathematics was driven by Hilbert's 23 problems and the development of physics. However, 20th Century Mathematics was mainly the golden age of applied mathematics, which is something Hilbert did not know. Mathematics, along with other sciences, is driven mainly by its times, the current state of the world, and probably not by people.

Modern geometry including differential geometry and topology quickly developed due to Einstein's theory of relativity (1905). The later developments of global differential geometry and algebraic geometry are related to string theory and gauge theory in physics as well. Algebraic geometry, famous because of its association with the Atiyah–Singer index theorem, also plays a central role in solving Fermat's last theorem in number theory. Its application in modern cryptography is called elliptical curve cryptography. Geometric analysis has been developing since the later part of the 20$^{th}$ Century and has become a branch of global differential geometry. It is believed that geometric analysis provided the tools for solving the Poincare conjecture. Modern algebra, on the other hand, was formulated in the 1930s by Van der Waerden in his book *Moderne Algebra*. Group theory was also developed in a very profound manner and included the completion of finite simple groups' classification.

Modern analysis developed mainly as a branch of mathematics called Functional Analysis, which also includes Harmonic Analysis. The base of this theory still consists of the variational and Dirichlet principles. The applications of these related theories were the primary focus of mathematics in the 20$^{th}$ century. Most of the numerical methods and optimization methods developed were related to modern analysis. On the other hand, probability and statistics were embedded in all aspects of everyday life including politics, economics, and finance. Game theory was developed, and stochastic processes began to play a central role in mathematical economics and risk analysis. The foundation of mathematics, which includes the axiomatic methods especially Hilbert's first and second problems, led to Godel's theory of incompleteness of arithmetic systems and the Turing machine in the 1930s. Such theoretical results were then developed to be combined with computing machinery.

In the late 1930 to 1940s, largely due to World War II, applied mathematics saw significant developments. C. Shannon founded information theory, Richard Hamming and Claude Shannon developed coding theory, and Norbert Wiener published *Cybernetics* for modern control theory. These theories are all related to probability and statistics. Soon after their development, operations research and systems theory emerged.

During the second half of the 20$^{th}$ Century, these new theories in mathematics had not developed in a solid manner, meaning that they lacked popular applications. Some of them are still in the stage of finding their "potential users" today. However, applied mathematics as a general subject has become enormously popular. Some of the most successful theories include:

- Functional analysis based on the variational method and Dirichlet principle for minimal surfaces and the foundation for the finite element method for partial differential equations in 1960-70.
- The simplex method for linear programming.
- Kalman filtering in control theory.
- Principle components method developed in 1901 but used later in image processing and other applications in statistics. This method used both linear algebra and numerical statistics.
- Algebraic applications in coding theory, especially the Reed-Solomon codes.
- Spline and Basic spline methods of function reconstruction methods in computer graphics, used widely in the automobile industry.
- Discrete algorithm design including graph-algorithms.
- Computational complexity based on the Turing machine.

We can also predict the potential development of the foundation for uncertainty mathematics. These theories include bifurcation theory, dynamical system theory, catastrophe theory, chaotic theory, and fractal geometry theory.

We hope that we have explained why Cook's theorem is one of the most important theorems in 20$^{th}$ Century mathematics. Cook's theory has a special meaning to mathematical logic since Cook's theory can be viewed as a branch of recursion theory. Freeman Dyson wrote a fantastic article in the 1980s, which recorded the story of how Godel insisted on his "Unfashionable pursuits" in logic. In fact, when (computer science) people were talking about the unfairness of the late appointment of professorship to Godel in the Institute of Advance Studies at Princeton, Stephen Cook reinstated the dignity of mathematical logic.

In mathematics, there are countless theories named after their inventors. So why do we not just call the NP-Completeness theory Cook's theory? This is because the NP-Complete theory has a very confusing name since most computer science students always think NP is referring to Non-Polynomial time algorithms. The NP-Complete theory should, rather, be called the P-Reduction theory. It is also true that Cook's theory can be called the Cook-Karp theory since Richard Karp took no delay in simplifying Cook's polynomial reduction based on Cook's theorem. In addition, Karp found 21 additional NP-hard problems.

In fact, Cook's theorem is also called the Cook-Levin theorem since Leonid Levin independently found the universal search method, which is equivalent to P-reduction. However, Levin did not use the NP class. Both Levin and Cook were very young when they made the break through in NP-Completeness almost forty years ago. Today's scientific research environments have changed significantly since late 1980s.

When Cook was asked what he would do if he submitted his paper today to one of the more selective conferences, such as the ACM SIG conferences, and his paper was rejected, he replied, "That is a good question." And immediately continued with, "I think I would submit to another place. If it is good work, it will be rewarded eventually!"

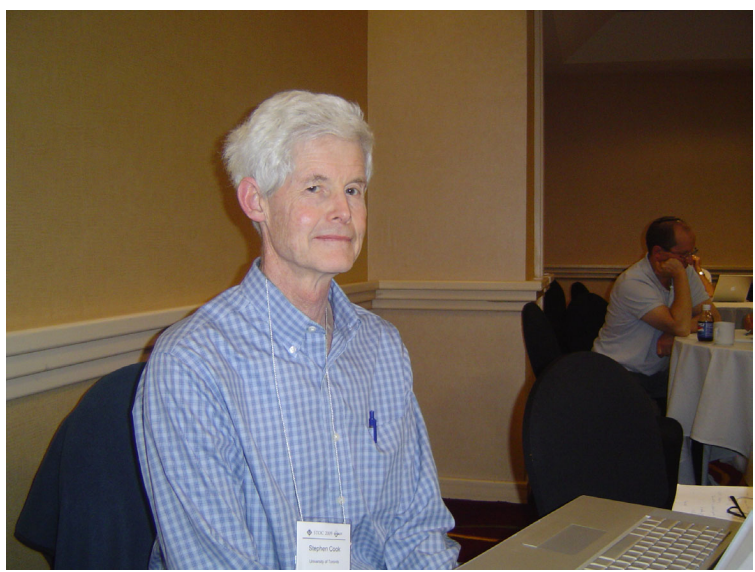